\documentclass[graybox]{svmult}

\usepackage{mathptmx}       
\usepackage{helvet}         
\usepackage{courier,mhchem}        
\usepackage{type1cm}        
\usepackage{makeidx}         
\usepackage{graphicx}        
\usepackage{multicol}        
\usepackage[bottom]{footmisc}

\usepackage{siunitx}
\usepackage{natbib}
\usepackage{amsmath,amssymb}
\usepackage{url}

\newcommand{\ecoli}{{\it E. coli}}

\makeindex             

\begin{document}

\title*{What is the `minimum inhibitory concentration' (MIC) of pexiganan acting on \textit{\textbf Escherichia coli}? --  A cautionary case study}
\titlerunning{Pexiganan action on \ecoli}
\author{Alys K Jepson, Jana Schwarz-Linek, Lloyd Ryan, Maxim G Ryadnov and Wilson C K Poon}
\authorrunning{Jepson et al.}
\institute{Alys Jepson, Jana Schwarz-Linek and Wilson Poon (w.poon@ed.ac.uk) \at SUPA and School of Physics \& Astronomy, The University of Edinburgh, Peter Guthrie Tait Road, Edinburgh EH9 3FD, Scotland, United Kingdom.\\Lloyd Ryan and Max Ryadnov \at National Physical Laboratory, Hampton Road, Teddington TW11 0LW, United Kingdom.
}
%
%
\maketitle

\abstract{We measured the minimum inhibitory concentration (MIC) of the antimicrobial peptide pexiganan acting on {\it Escherichia coli}, and report an intrinsic variability in such measurements. These results led to a detailed study of the effect of pexiganan on the growth curve of \ecoli, using a plate reader and manual plating (i.e. time-kill curves). The measured growth curves, together with single-cell observations and peptide depletion assays, suggested that addition of a sub-MIC concentration of pexiganan to a population of this bacterium killed a fraction of the cells, reducing peptide activity during the process, while leaving the remaining cells unaffected. This pharmacodynamic hypothesis suggests a considerable inoculum effect, which we quantified. Our results cast doubt on the use of the MIC as `a measure of the concentration needed for peptide action' and show how `coarse-grained' studies at the population level give vital information for the correct planning and interpretation of MIC measurements. }

\section{Introduction}
\label{sec:intro}

The discovery of the $\beta$-lactam antibiotic penicillin by Fleming in 1928 was a milestone of $20^{\rm th}$-century medicine. Today, the rampant spread of antimicrobial resistance (AMR) constitutes a grand challenge facing medical science in the new century \citep{Aminov2010}, which, if not met, will turn a `strep throat' back into a life threatening illness. To confront AMR, more effective ways of using of existing agents, including dosing regimes less prone to generating AMR, are urgently needed, as are new agents. In either case, standardised measures of effectiveness allowing science-based comparison between different agents are clearly required. 

In the global effort to confront AMR, antimicrobial peptides (AMPs) have attracted considerable attention \citep{Schneider2012}. These short peptides ($\approx 10$ to 50 residues) are widely distributed among metazoans, with diverse sequence and structure. Certain motifs recur, e.g. high net positive charge under physiological conditions, and/or $\alpha$-helical conformation in solution or upon binding to membranes; but these motifs are not universal. AMPs are effective against a wide spectrum of bacteria, viruses and fungi {\it in natura}. The hope is that some natural or synthetic AMPs may be suitable therapeutic antimicrobial agents, especially against AMR strains. 

The 23-residue AMP magainin-2 secreted by the African clawed frog {\it Xenopus laevis} and its `relatives' have attracted particular attention. Pexiganan (or MSI-78), a synthetic 22-residue magainin analogue \citep{Gottler2009}, was trialed for topical treatment of diabetic foot ulcers, but was denied approval in 1999 because it seemed no more effective than antibiotics already in use for such ulcers \citep{Moore2003}; however, future clinical approval remains a possibility \citep{Gottler2009}. Partly due to on-going efforts to secure such approval, pexiganan has been well studied. 

A large biophysical literature exists on AMPs in general, and pexiganan in particular, focussing on the molecular {\it modus operandi}. Partly as a result of substantial research into how pexiganan and similar AMPs interact with lipid bilayers in unilamellar vesicles \citep{Gottler2009}, it is widely believed that this and other $\alpha$-helical AMPs lyse bacteria by membrane poration. 

We consider the other end of the length scale spectrum, and report a study of the {\it modus operandi} of pexiganan on \ecoli\, at the population level. Such `coarse-grained' studies using rather classical methods (albeit in updated, high-throughput, forms) are seldom performed today. Our results show how such work is needed to complement molecular-level studies,  preventing the misinterpretation of pharmacodynamic measurements performed to judge the concentration required for antimicrobial action against live bacteria. 

We start by measuring the minimum inhibitory concentration (MIC), which is the most important `one-number characterisation' of the effectiveness of an antimicrobial agent against a target organism. Loosely, it is the minimum concentration of an antimicrobial agent necessary to cause stasis (no growth). We re-determine the MIC of pexiganan on \ecoli, but using more repeat experiments than has ever been reported before. A critique of using MIC to characterise potency based on our measurements leads us to study the effect of pexiganan on the growth curve of \ecoli, which turns out to be strikingly different from the way many classical antibiotics change the growth curve of the same bacterium.  

Our growth curves, along with single-cell observations, peptide depletion assays and time-kill curves, suggest that adding sub-MIC concentrations of this AMP to a population of \ecoli\, rapidly kills a fraction of the cells, leaving the rest to grow unaffected while at the same time removing active AMP molecules from the medium. As \cite{Udekwu2009} have previously suggested, the depletion or deactivation of an antibiotic causes an `inoculum effect' (dependence of MIC on initial inoculum concentration), which we quantify for pexiganan. We end by discussing the implications of our findings on the interpretation of the MIC in mechanistic and pharmacodynamic contexts.

\section{Materials and methods}
\label{sec:blah}

\subsection{Bacteria culture}

We worked with \ecoli\, K-12 derived strain MG1655 \citep{coligenome}. Five colonies, grown on an Mueller Hinton Broth (MHB) agar plate, were touched with a sterile loop and introduced to a 5 ml MHB liquid culture, which were grown at $37^{\circ}$C and 200rpm to OD=0.5 (600nm). 
We also grew bacteria in the filtered supernatant of cells lysed by sonication. Ten 30 second pulses of sonication applied to \SI{10}{\milli\litre} of \ecoli\ culture resting on ice achieved a 99.98\% reduction in viability. The supernatant was filtered ($0.22\mu$m) to remove the surviving cells. 

\subsection{Pexiganan}

Pexiganan (GIGKFLKKAKKFGKAFVKILKK-\ce{NH2}) was synthesised on a Liberty microwave peptide synthesizer (CEM Corporation) using standard solid phase Fmoc protocols on Rink amide-MBHA resins with HCTU/DIPEA as coupling reagents. Peptides were purified by semi-preparative RP-HPLC on a JASCO HPLC system (model PU-980, Tokyo, Japan) and confirmed by MALDI-ToF mass spectrometry (Bruker Daltonics Ltd, UK) with $\alpha$-cyano-4-hydroxycinnamic acid as the matrix. 

We prepared stock solutions at 2mM in sterile, distilled water, which were stored at $-20^{\circ}$C and defrosted immediately before using and refreezing.

\subsection{Growth curves and MIC}

We followed a published protocol for MIC determination using microdilution assays in microtiter plates \citep{Wiegand2008} that is consistent with the guidelines of the Clinical and Laboratory Standards Institute and the European Committee on Antimicrobial Susceptibility testing. MIC assays were prepared in 96-well polystyrene microtiter plates (Greiner) with \SI{200}{\micro\litre} cylindrical wells. Initial inoculum sizes of $n_0 = 5 \times 10^5$ \si{\mbox{cell}\per\milli\litre} were incubated in and their optical density read (at \SI{600}{\nano\meter}) by a FLUOstar Optima (BMG Labtech) plate reader with lid-covered plates that allowed air flow. We checked that our MIC results were the same using either polystyrene or polypropylene plates. 
Note that we report pexiganan concentrations in \si{\micro\mbox{M}} to facilitate comparison with cell concentrations in \si{\mbox{cells}\per\milli\litre}, where $\SI{0.4}{\micro\mbox{M}} = \SI{1}{\micro\gram}$/\si{\milli\litre} (using the molecular weight  of \SI{2477}{\gram\per\mol}). Literature values in \si{\micro\gram}/\si{\milli\litre} have been converted to and quoted in \si{\micro\mbox{M}}.

\subsection{Single-cell imaging}

To determine times to first division, we took time-lapsed phase-contrast images of cells using a Nikon TE300 Eclipse inverted microscope with a $100\times$ PH3 oil-immersion objective and a CoolSNAP HG$^2$ CCD camera (Photometrics). MHB agar pads were set into adhesive Gene Frames (Thermo Scientific). Pexiganan at $3\mu$M was added to \ecoli\ at OD=0.5 and the solution was left to incubate for 3 minutes before pipetting \SI{1}{\micro\litre} onto the agar pad. The \ecoli\ were spread by tipping the microscope slide and within $\sim4.5$ minutes all liquid had, by eye, disappeared. We then mounted a glass coverslip to the gene frame in contact with the agar pad. The sample was immediately transferred to the microscope, pre-heated to $37^{\circ}$C in a temperature-controlled box, for observation. 

\subsection{Time-kill curves}
\label{sec:TKC}

We grew \ecoli\, MG1655 following the same protocol as for MIC assays and diluted to $5 \times 10^5$ \si{\mbox{cell}\per\milli\litre}. We worked with \SI{3}{\milli\litre} of suspension in two tubes (Greiner, polystyrene, 50ml). Pexiganan was added to one of these at time $t = 0$ min and both incubated at \SI{37}{\celsius} and shaken at 200~rpm. A \SI{100}{\micro\litre} sample was removed from both tubes \SI{1}{\minute} after peptide addition, a range of ten-fold dilutions were spread onto MHB agar plates in triplicate. No more than 5 samples were removed, resulting in a 17\% volume reduction, which is somewhat above the recommended maximum reduction in standard protocols. The agar plates were incubated at \SI{30}{\celsius} for \SI{16}{\hour} before the colonies were counted manually and density of cells calculated in cell-forming units (CFU) per ml. 

\subsection{Pharmacodynamic studies}\label{sec:pharma}

We followed literature procedures \citep{Udekwu2009} to determine the effect of residual AMP. Two tubes containing 5ml of MHB and $40\mu$M of pexiganan were prepared, one of which was inoculated with $5\times10^6$ cells/ml. After 18 h incubation both suspensions were filtered and the supernatant used to set up two MIC assays in a 96 well plate each, alongside two replicate control MIC assays using the standard protocol. A further two assays were set up with supernatant from the tube which had contained both cells and peptide, with $20\mu$M of freshly-added pexiganan.

\section{Minimum inhibitory concentration}
\label{sec:MIC}

\begin{figure}[t]
\begin{center}
\includegraphics[width=0.48\textwidth]{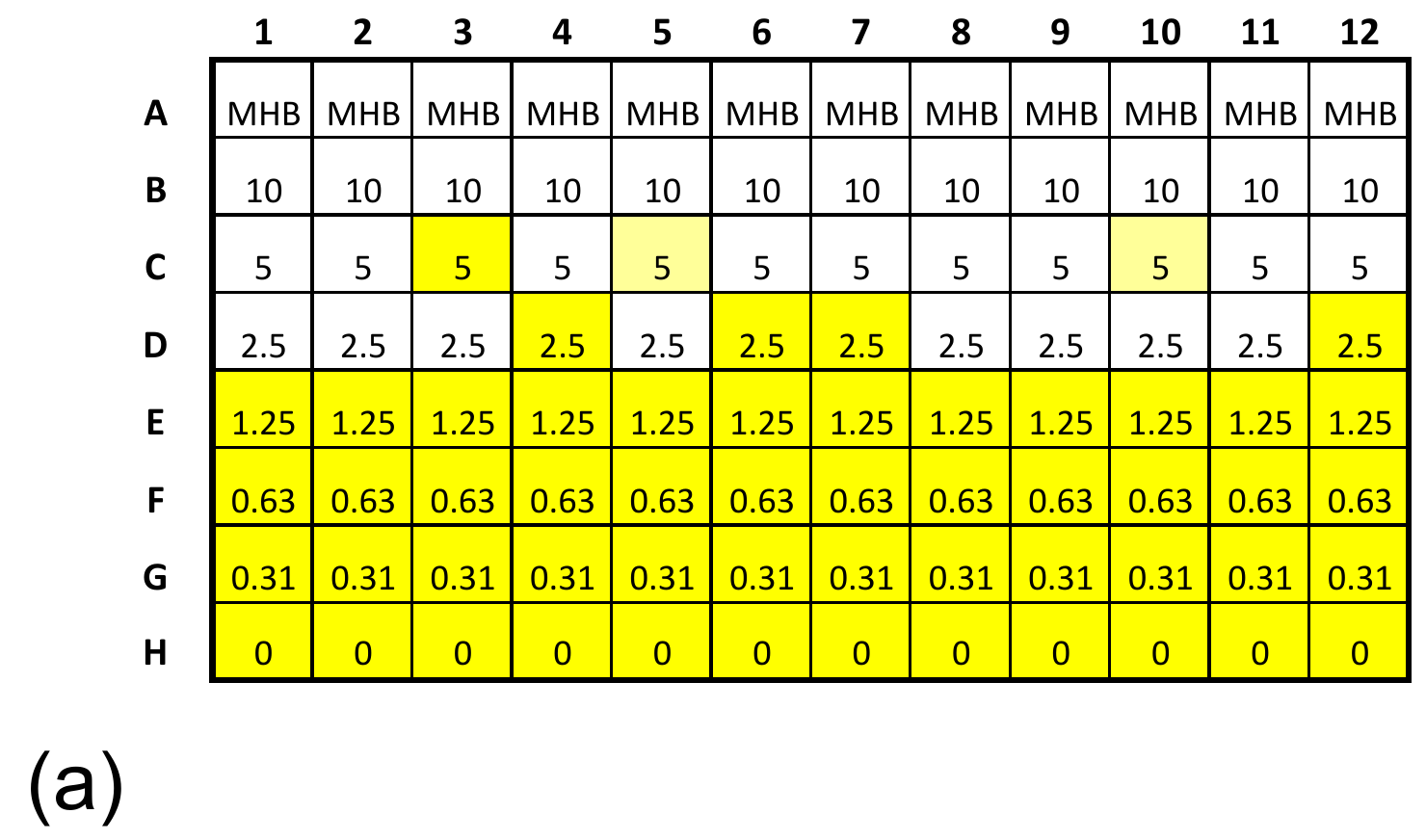}
\includegraphics[width=0.48\textwidth]{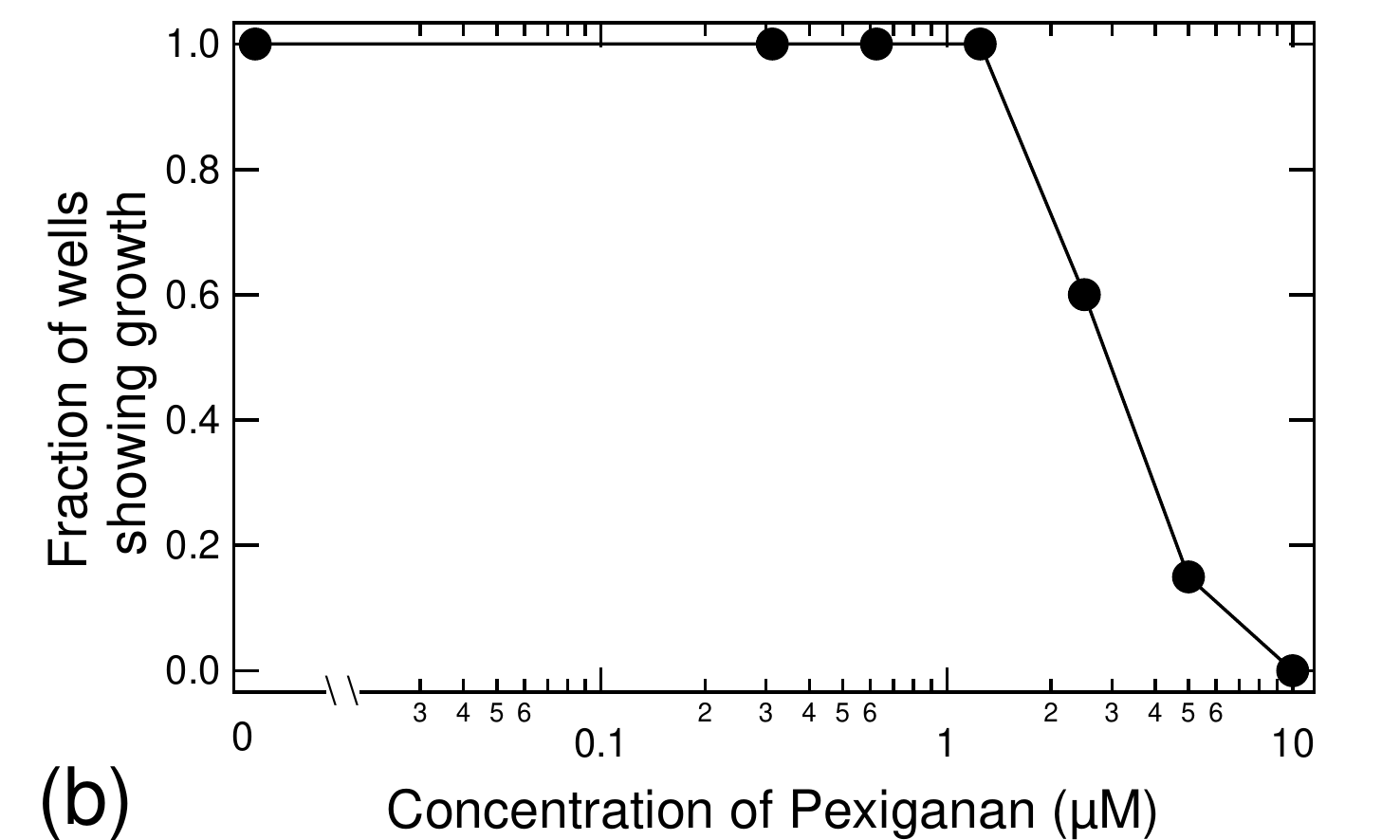}
\end{center}
\caption{(a) Data from one dilution experiment in a 96-well plate. Numbers are pexiganan concentrations in \si{\micro\mbox{M}} in the \SI{200}{\micro\litre} wells, each inoculated with $n_0 = 5\times 10^5\,\si{\mbox{cells}\per\milli\litre}$. MHB = buffer with no cells. Each column is one dilution series. Yellow =  wells with an OD $\approx 10\times$ the OD of MHB wells after \SI{24}{\hour}. (b) Fraction of wells that showed apparent bacterial growth for 20 replicates as a function of pexiganan concentration, with the zero pexiganan point plotted on the left end of the logarithmic horizontal axis.}
\label{fig:96well}       
\end{figure}

Figure~\ref{fig:96well}(a) shows representative raw data for our determination of the MIC of pexiganan for \ecoli\, MG1655 using microdilution assays in microtiter plates, with growth after \SI{24}{\hour} detected by a plate reader.\footnote{See \url{http://datashare.is.ed.ac.uk/handle/10283/1885}  to access relevant data on which this article is based.} In our 96-well plate, each of the 12 columns constitutes a separate MIC determination. Thus, e.g., in column~4 of the plate shown, the minimum concentrating showing no growth at \SI{24}{\hour} is \SI{5}{\micro\mbox{M}}, which therefore by definition is the MIC from this particular dilution series. In these 12 replicates, then, five return a MIC of \SI{2.5}{\micro\mbox{M}} (columns 1, 2, 8, 9 and 11), four return a MIC of \SI{5}{\micro\mbox{M}} (columns 4, 6, 7 and 12), and three columns (3, 5 and 10) show a seemingly `impossible' pattern of `re-entrant growth': after being inhibited at \SI{2.5}{\micro\mbox{M}}, growth apparently restarted at \SI{5}{\micro\mbox{M}}.\footnote{Continuing the experiment to \SI{72}{\hour} did not change the observed growth/no-growth pattern.} 

What, then, is {\it the} MIC of pexiganan acting on \ecoli\, MG1655? Our data, Figure~\ref{fig:96well}(a), do not allow us to assign a unique value, but if such a value exists, then it lies in the region of 2.5 to \SI{5}{\micro\mbox{M}}. Previous studies have returned values in the range of 3.2 to \SI{12.8}{\micro\mbox{M}} against various isolates \citep{Fuchs1998,Ge1999}, placing our range of \SI{2.5}{\micro\mbox{M}} to \SI{5}{\micro\mbox{M}} at the low end of the spectrum.\footnote{This is perhaps unsurprising given that MG1655 is a laboratory strain that has been described as `deceitful delinquents growing old disgracefully' \citep{Hobman}.}

However, to compare our results and previous work in terms of a range of values masks a fundamental qualitative difference between our work and almost all previous MIC measurements. We, like the majority of the literature, follow  best-practice guidelines \citep{Wiegand2008,CLSI}. The difference is that we perform multiple (12) replicates of the same MIC assay on a single plate. This is almost never done in the literature. Instead, commercial `MIC plates' are used with wells pre-loaded to give single dilution series of multiple antimicrobial agents. Alternatively, the same dilution series of a single agent is used to test multiple organisms, again with one dilution series per organism.\footnote{E.g., no replicats are suggested in the guidelines for loading a 96-well plate for testing a range of agents on \ecoli\, and {\it Salmonella} isolates in a WHO project \citep{SalmSurv}.} The range of literature values quoted above for pexiganan acting on \ecoli\, therefore arises from single-dilution-series measurements on many isolates, whereas our own range of values arises from variability between multiple replicates on a single strain, Figure~\ref{fig:96well}(a). 

When a single dilution series is used, as is the case in the majority literature, the issue of disagreement between replicates does not arise. On the rare occasion where two replicates are prepared and they disagree, such as our columns 7 and 8 (Figure~\ref{fig:96well}), the disagreement is typically ascribed to `dilution error'. Literature protocols do occasionally mention `re-entrant growth', but would ascribe this to  accidental `skip' or to `single well contamination'. In particular, if we follow \cite{SalmSurv}, we should identify our wells D3, D5, D10 as `skips', and in these cases take the `true MIC' to be \SI{10}{\micro\mbox{M}}. \footnote{\sloppy See Figure~2 in \cite{SalmSurv}. If, alternatively, we identify our wells C3, C5 and C10, Figure~\ref{fig:96well}, as contaminated, then the `true MIC' would be taken to be \SI{2.5}{\micro\mbox{M}}.}
Repeated dilution error, skip or contamination at the same point of our multiple dilution series seem highly improbable. Moreover, repeated measurements showed similar patterns of variability seen in Figure~\ref{fig:96well}(a), including `reentrant growth'. 
Thus, the variability revealed by our results is intrinsic, and reporting a single MIC is misleading. A better way to summarise our findings is to plot the fraction of wells showing growth as a function of pexiganan concentration. Figure~\ref{fig:96well}(b) shows such a plot for the 12 replicates shown in Figure~\ref{fig:96well}(a) and another 8 replicates performed using the same peptide stock and inoculum. 

Intrinsic variability is consistent with a previous meta-study \citep{Annis2005}, which ascribed half of the variability uncovered in a survey of literature values of the MICs of various antibiotics against \ecoli\, and {\it Staphylococcus aureus} to laboratory-to-laboratory differences. Presumably, then, it is possible that at least part of the other half of the variability is attributable to intrinsic causes, although Annis and Craig attributed this `commonly shown 3-fold dilution range' entirely to environmental factors such as temperature, inoculum size and incubation time.

To begin to elucidate the source of the MIC variability, we turn to consider the full growth curves that were collected and used to generate the MIC data.

\section{Sub-MIC growth curves}

\begin{figure}[t]
\begin{center}
\includegraphics[width=0.95\textwidth]{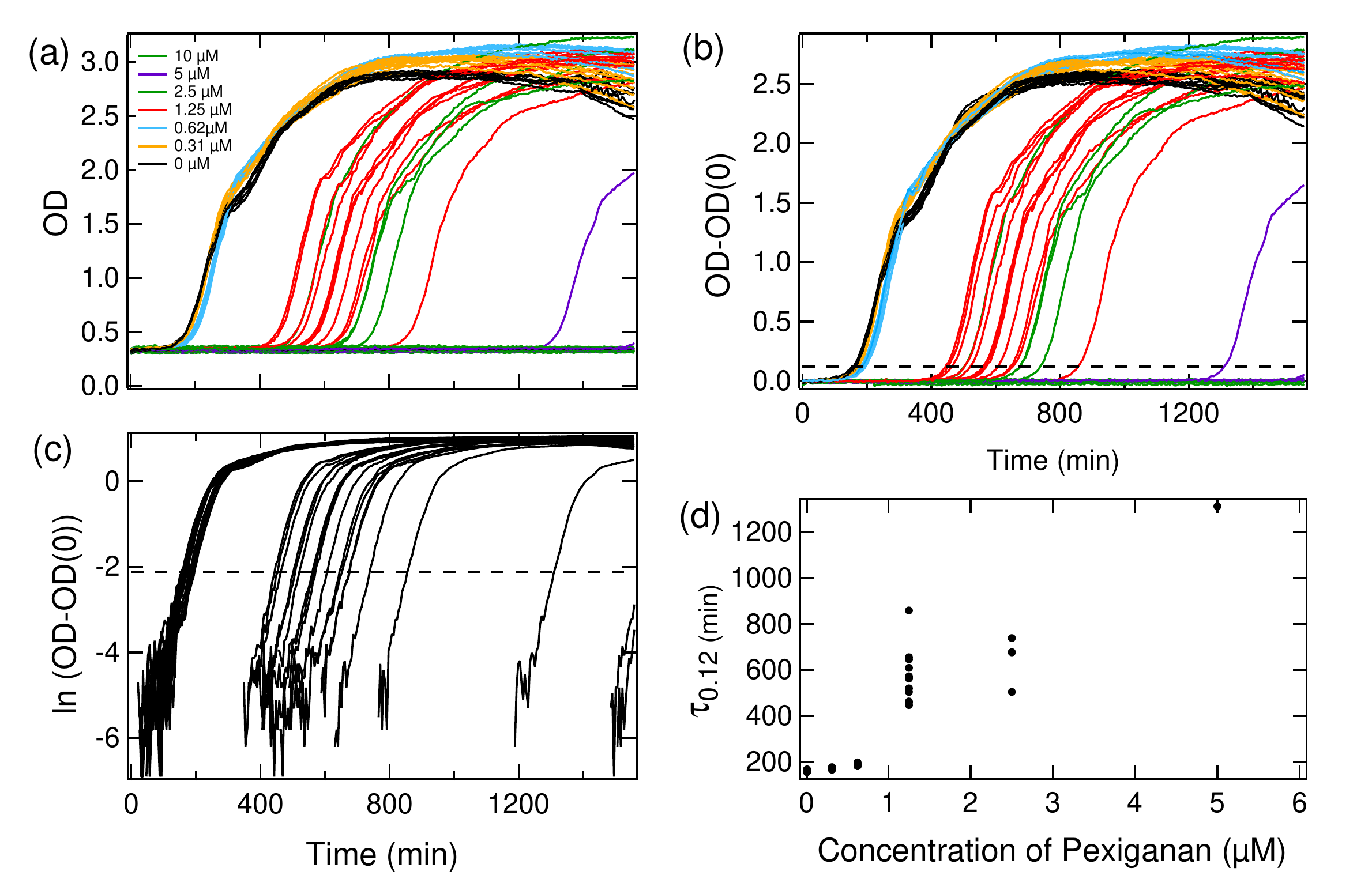}
\end{center}
\caption{Sub-MIC growth curves for \ecoli\, with an initial inoculum size of $n_0 = 5\times 10^5\,\si{\mbox{cells}\per\milli\litre}$ at the same pexiganan concentrations used for the 96-well plate experiment in Figure~\ref{fig:96well} and stated in the legend. (a) Raw OD data against time. (b) The same data with OD at $t=0$ subtracted. (c) Log-linear plot of the data in (b). (d) The time to reach OD = 0.12, $\tau_{0.12}$, plotted against concentration of pexiganan. In parts (b) and (c), OD = 10.12 is shown as dashed lines.}
\label{fig:subMIC}       
\end{figure}

Figure~\ref{fig:subMIC}(a) shows a typical set of growth curves  of \ecoli\ MG1655 at sub-MIC concentrations of pexiganan (background subtracted in Figure~\ref{fig:subMIC}(b) and (c)), from the MIC assays displayed in Figure~\ref{fig:96well}(a). While such curves are collected in every microtiter assay, they are seldom, if ever, presented or interpreted. Exceptions concern the response of \ecoli\, to tetracycline and amoxicillin \citep{Schuurmans2009} and to cefotaxime \citep{Baraban2011}. At sub-MIC concentrations tetracycline and cefotaxime reduce the population growth rate and stationary level whilst amoxicillin does not significantly affect the growth curves until the MIC is reached.   

Pexiganan does not influence the maximum growth rate (measured to be $\alpha = 0.034\pm0.01$min$^{-1}$) or the stationary level of the population (a background subtracted OD of $\approx 2.5$), but lengthens the time until growth is detected\footnote{We detected growth in a well at $2\times10^{7}$ cells/ml, similar to what was reported previously for multi-well plate readers \citep{Pin2006,Metris2006}. Lengthened detection times due to peptide action recalls the `virtual colony count' (VCC) approach developed to measure defensin activity \citep{Ericksen2005}. However, unlike in VCC, bacteria in our case were exposed to peptides in their growth medium rather than grown in a peptide-free medium after exposure.} in a concentration dependent manner. Figure \ref{fig:subMIC}(c) plots the time taken to measure an OD of 0.12 above the OD of MHB ($\tau_{0.12}$), which increases with pexiganan concentration. Note that without pexiganan, the population takes $160\pm10$ min to grow to OD = 0.12 =~$1\pm 0.4\times10^8$ cell/ml. An initial inoculum of $5 \times 10^5$~cells/ml will reach $1\times10^8$~cell/ml in $\alpha^{-1} \ln (2\times 10^2) = \SI{156}{\minute}$ with our measured $\alpha$. The population lag time is therefore immeasurably small.

The variation in $\tau_{0.12}$ between wells increases with pexiganan concentration. In particular, the large differences in $\tau_{0.12}$ between replicates at $2.5\mu$M correlates well with the observation, Figure~\ref{fig:96well}, that some wells do not show growth within 24 h at this concentration. No replicate experiments were reported in previous growth curve studies. In the case of cefotaxime \citep{Baraban2011}, the smooth variation of growth curves at closely-spaced antibiotic concentrations allows us to conclude that there should be little variation between replicates. We therefore infer an intriguing difference vis-\`a-vis stocasticity between pexiganan and cefotaxime. We cannot draw a similar conclusion from data for tetracycline and amoxicillin \citep{Schuurmans2009} because only a few concentrations were studied.

\section{Single cell observations}

\begin{figure}[t]
\begin{center}
\includegraphics[width=0.6\textwidth]{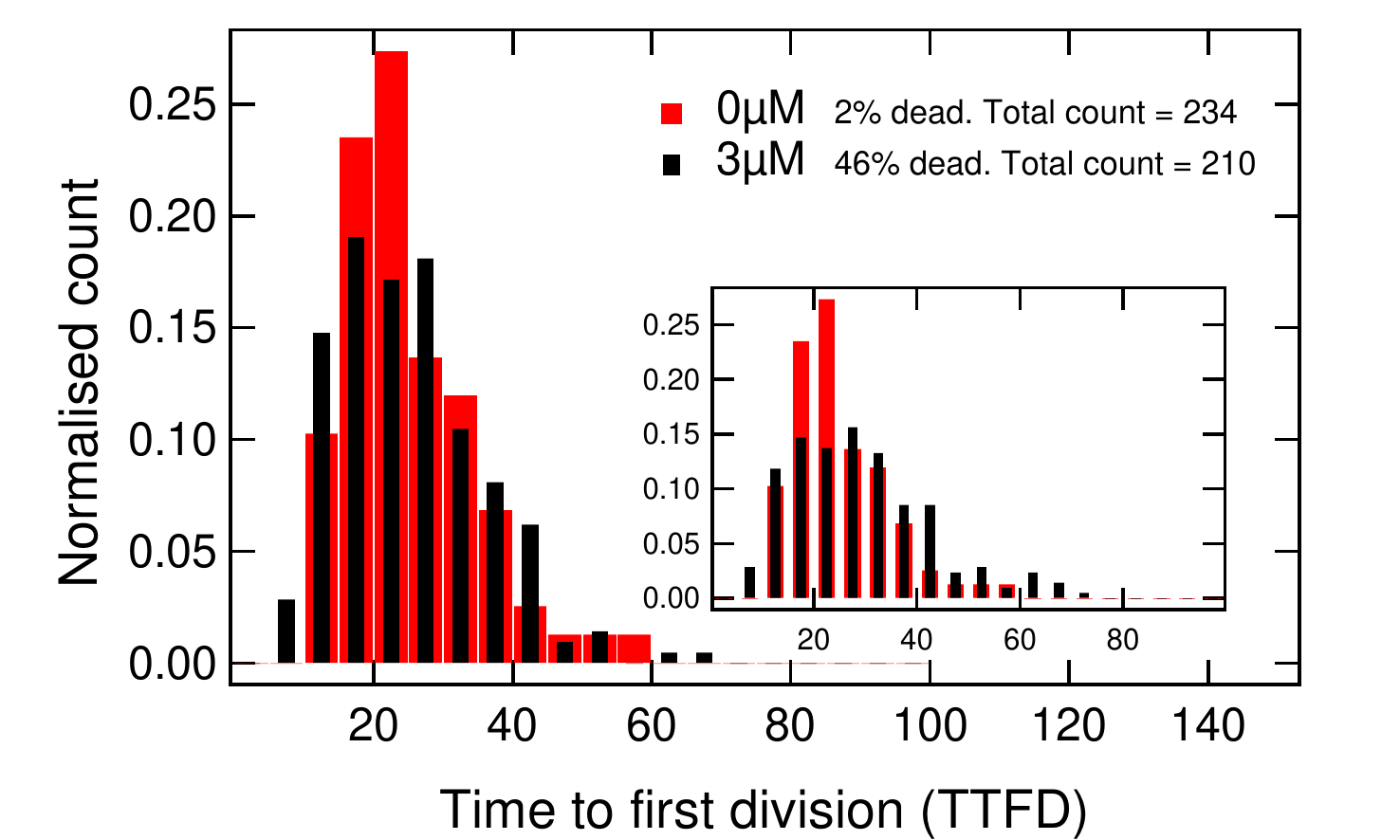}
\end{center}
\caption{Normalised histogram of times to first division for \ecoli\ on agar. Unexposed to pexiganan (red) and exposed to $3\mu$M pexiganan for 3 min (black) Inset shows the same plot if the time to second division is counted for all first divisions resulting in the death of a daughter cell.}
\label{fig:SC}       
\end{figure}

Next, we probe the cause of the observed effect of pexiganan concentration on $\tau_{0.12}$, the growth curve detection time. There are two generic ways in which the peptide could affect cells at early times to increase $\tau_{0.12}$. Either some of the initial inoculum dies, and the remaining cells take longer to grow to a given density even if their growth rate remains unaltered, or growth of all the cells is retarded, giving the same macroscopic observed effect on $\tau_{0.12}$. 
To observe directly the early-time effect of sub-MIC concentrations of pexiganan, we imaged cells spread on agar and recorded the time to first division (TTFD) of each cell, thought to be the sum of the lag time and first generation time \citep{Rasch2007, Metris2005}. We chose conditions (caption, Figure~\ref{fig:SC}) that resulted in sufficient cell death to demonstrate the effect of pexiganan, whilst leaving enough live cells to collect meaningful data.

The TTFD distribution for the pexiganan-free control, Figure~\ref{fig:SC} (red), is spread over $\sim~\SI{50}{\minute}$ due to heterogeneities in growth stages, single-cell lag times\footnote{ Cells with the shortest single cell lag times dominate the population lag \citep{Baranyi1998}, which can be crudely calculated to be $\sim10$ min after the transfer procedure from liquid culture to agar.} and generation times. On exposure to $3\mu$M of pexiganan for 3 min before being placed on the agar, a fraction of the cells never go on to divide. Some are clearly dead (showing less contrast and do not grow) by the time the first image was taken, but others grow initially and then stop at later times. Some cells lyse suddenly whilst other fade over time. Interestingly, 29\% of the cells which grow and divide give rise to a daughter cell that subsequently dies. Importantly, the TTFD distribution of cells that survive and divide to form colonies does not seem to differ greatly to that of the control sample, Figure~\ref{fig:SC} (black and red respectively). Including cells whose daughters subsequently die shifts the distribution only slightly (inset). 

These data suggest that exposure to pexiganan results either in death or leaves the cells unaffected to go on to grow and divide, at least for cells inoculated onto agar. A number of caveats are, however, in order. First, the concentration of peptides the cells are exposed to is uncontrolled while the inoculum drop is evaporating after being placed on the agar surface. Thereafter, the peptides are be free to diffuse into the agar away from the cells. It is therefore possible that under more prolonged exposure than is possible under our experimental conditions, e.g., in liquid medium, the growth of surviving cells would be retarded. Nevertheless, our experiments certainly show that the main effect of pexiganan is to kill a proportion of the cells rapidly.

\section{Depletion of active peptide}

It is important to know whether this rapid initial killing of cells significantly affects the concentration of free peptides. Towards this end, we first determine if the peptide is depleted with time in our assays we performed the MIC experiments detailed in section \ref{sec:pharma}. MIC assays using pre-incubated peptide showed that the activity of the peptide degrades by a factor of 2 to 4 times over 24 h in MHB at $37^{\circ}$C, perhaps due to peptide aggregation, common for magainins, or adhesion to components of the medium. However, this is a small contribution to what was found when bacteria had also been in the solution. An assay containing $40\mu$M of pexiganan and $5\times10^6$ cell/ml, after being incubated and then filtered to remove any survivors, no longer had any measurable affect on the growth of a new inoculum, so that (from Section~\ref{sec:MIC}) we know that the pexiganan concentration must be below \SI{2.5}{\micro\mbox{M}}. In other words, the activity of the peptide was reduced more than 16-fold. When an additional $20\mu$M of peptide was added to this solution and serially diluted with MHB it resulted in a MIC $4\times$ greater that that of a control. Not only had the original peptide been depleted but the additional peptide showed less activity in this solution than in MHB. Similar experiments conducted without filtering gave the same results, so that the filtering was not responsible for removing active peptide molecules or complexes.

Following these findings, we used a suspension containing the filtered remains of sonicated cells grown in MHB to prepare a MIC assay instead of MHB alone. Even at the highest concentration of $20\mu$M pexiganan, there was no measurable effect on the inoculum of $5\times10^5$ cell/ml, suggesting a reduction in the antimicrobial action of the peptide of by 10 to 20 fold. It appears that the presence of lysed bacteria reduces peptide activity. A plausible hypothesis is that positively-charged peptides are depleted by adhesion to negatively-charged DNA and proteins, and/or by the action of proteolytic enzymes from lysed cells.

\section{Sub-MIC time-kill curves}

\begin{figure}[t]
\begin{center}
\includegraphics[width=0.6\textwidth]{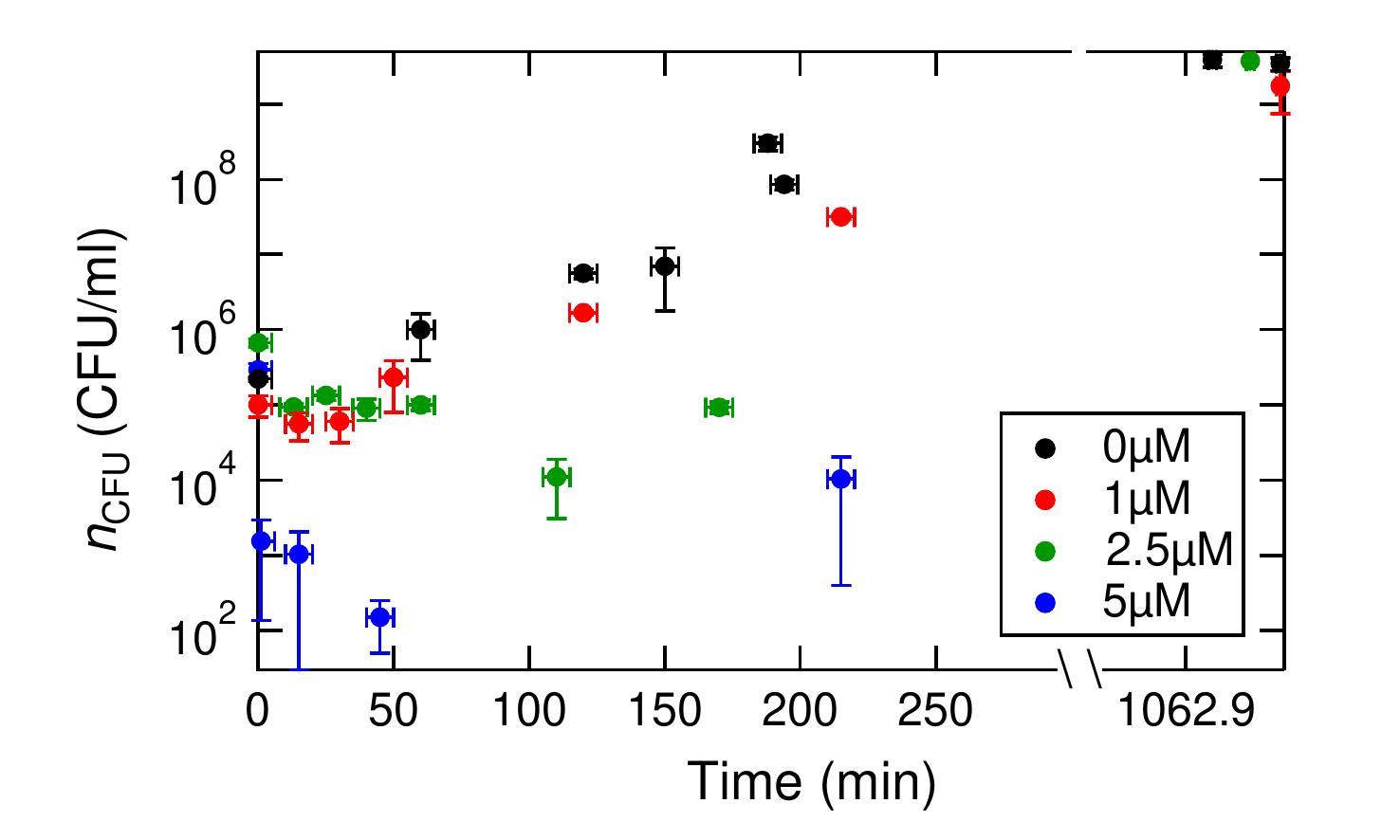}
\end{center}
\caption{Platecounted time-kill curves in MHB for pexiganan concentrations shown in the legend for the inoculum size $n_0 = 5\times 10^5\,\si{\mbox{cells}\per\milli\litre}$.} 
\label{fig:KC_MHB}       
\end{figure}

If we are right that pexiganan kills a fraction of cells at short times upon addition, our growth curves ought to show an initial decrease in numbers. This is not borne out by our plate reader data, Figure~\ref{fig:subMIC}. We now demonstrate that this is simply because of lack of detection sensitivity. By using plate counts to measure the number of viable \ecoli\ as a function of time after the addition of pexiganan to their medium, i.e. by determining time-kill curves (Section~\ref{sec:TKC}), it is possible to detect much lower cell densities than is possible using a plate reader. We added pexiganan at $t=\SI{0}{\minute}$ to concentrations of 1, 2.5, 5 or \SI{10}{\micro\mbox{M}} and collected plate-count data for the first 200 min, and then once the following day.

Figure \ref{fig:KC_MHB}(a) shows time kill curves for an inoculum size of $5\times10^5$ cell/ml, the same as that used for MIC assays in the plate reader. As can be expected from the results shown in Section~\ref{sec:MIC}, no viable cells were recorded at any time point at \SI{10}{\micro\mbox{M}}. At lower concentrations the density of viable cells in the suspension drops at early times; the higher the peptide concentration, the larger the drop in initial viability. This is as expected from the only previous publication of pexiganan time-kill curves \citep{Ge1999}. These measurements however were performed at above-MIC concentrations, and only monitored for 120 minutes. 

We measured at lower concentrations and for longer and therefore observe the viability of \ecoli\ increasing again at later times.  The higher the peptide concentration, the longer before recovery starts; however, after one day, all samples have recovered to the same level. 

Regrowth after an initial drop in viability has been reported in the time-kill curves of other AMPs \citep{Matsuzaki1998,McGrath2013,Spindler2011}. It has been attributed to `resistance', with little evidence given in support. To verify that no resistance had developed in our case, we exposed \ecoli\ that had regrown in our experiments to 2.5 and $5\mu$M pexiganan, and found them as susceptible as cells from the parent population.

Our results confirm that the lack of an initial phase of negative growth in our plate reader data, Figure~\ref{fig:subMIC}, is simply due to lack of detection sensitivity, and strengthens the case for our suggestion that pexiganan kills a fraction of the cells at early times, leaving the rest to grow unaffected. The literature has only recently suggested that such non-monotonic time-kill curves can be due to a dynamic state of balanced division and death rather than the historically assumed sub-population of non-dividing persister cells \citep{Wakamoto2013}.

The killing of a fraction of cells in a clonal population leaving the rest of the cells unaffected is striking. Given the large number of peptides per bacterium in the system ($\sim 10^9$, much more than is needed to cover each cell in a monolayer however the peptide is oriented) \citep{Wimley2010}, the observed heterogeneity is unlikely to be due to fluctuations in the number of peptides strongly interacting with each cell. Instead, the heterogeneity most likely arises from phenotypic variations in a property or properties of single cells. Our evidence shows that the survivors are not a sub-population of non-dividing cells. Perhaps, then, cell surface heterogeneities, \emph{e.g.}  in the structure of the lipopolysaccharides \citep{Lerouge2002} or in the expression of fimbrae \citep{Abraham1985}, are responsible for our observations.

Note that at peptide concentrations approaching the MIC, e.g., $\gtrsim \SI{2.5}{\micro\mbox{M}}$ for the inoculum size relevant to Figure~\ref{fig:KC_MHB}, the number of viable cells drops to rather low levels (the minima at \SI{2.5}{\micro\mbox{M}} and \SI{5}{\micro\mbox{M}} in Figure~\ref{fig:KC_MHB} correspond to 2000 and 20 cells per well respectively) before net growth begins. Such low numbers lead to large fluctuations; in particular, some replicates could easily have all cells eradicated at these peptide concentrations. This is the source of the observed fluctuations in Figure~\ref{fig:96well}, which make it difficult to determine `the MIC' based on a single measurement. 

\section{The inoculum effect}\label{section:IE}

If pexiganan kills a certain fraction of cells outright, depletes the number of active peptides, and leaves the remaining cells to grow, then there should be a strong inoculum effect (IE): the MIC should be higher for higher inoculum densities. The IE is important because the {\it in vitro} population density of pathogens could be significantly higher than typical inocula cell densities in MIC assays. The magnitude of the IE for a small number of inocula concentrations is reported in the antimicrobial literature \citep{Udekwu2009}, including for some AMPs \citep{Levison1993,Jones1994}; but possible causative mechanisms have seldom been discussed, partly because data for MICs over a large range of inocula densities are, with very few exceptions \citep{Udekwu2009}, not reported. 

\begin{figure}[t]
\begin{center}
\includegraphics[width=0.49\textwidth]{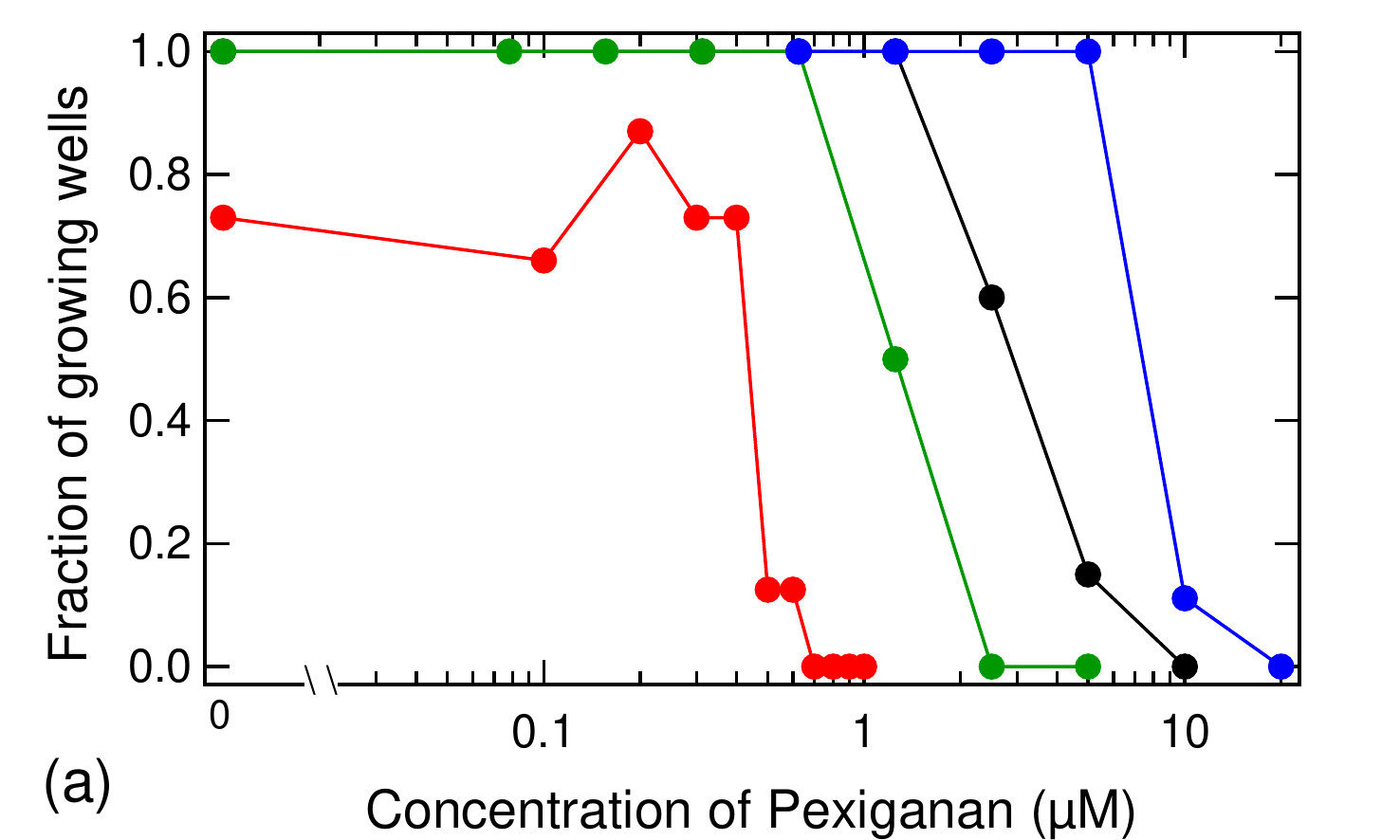}
\includegraphics[width=0.498\textwidth]{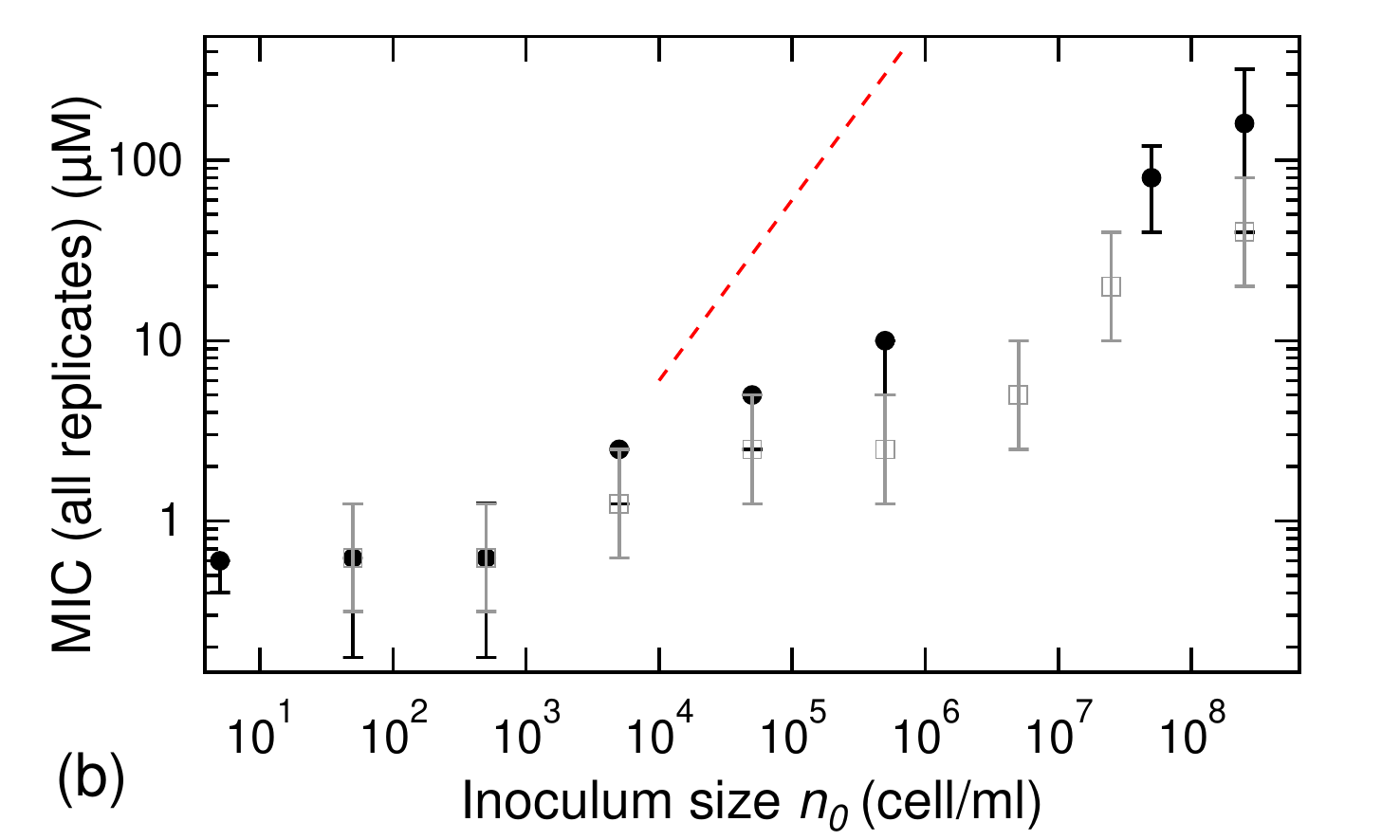}
\end{center}
\caption{(a) Fraction of growing wells as a function of pexiganan concentration, for inocula of $n_0 = 2.5\times 10^7$ (blue), $n_0 = 5\times 10^5$ (black), $n_0 = 5\times 10^3$ (green) and $n_0 = 5 ~\si{\mbox{cells}\per\milli\litre}\,(\equiv 1 \mbox{cell/well})$ (red). (b) The MIC, defined as the concentration of pexiganan at which no replicate grows, plotted agains inoculum size, $n_0$, for one set of experiments using the same pexiganan stock (grey) and data collated from many tests (black). The dashed line has unit slope.}
\label{fig:4N0}       
\end{figure}

We report such data for pexiganan acting on \ecoli. Figure \ref{fig:4N0}(a) shows the fraction of growing wells against pexiganan concentration for four different inoculum sizes, $n_0 = 2.5\times10^7$, $5\times10^5$ (our standard inoculum size), $5\times10^3$ and 5 cell/ml.\footnote{Note that 5~cells/ml $\equiv$ 1~cell/well, so that the fraction of growing wells is $<1$ at zero pexiganan.} Qualitatively, these data are similar to those shown in Figure~\ref{fig:subMIC}. Quantitatively, the pexiganan concentration at which the fraction of growing wells drops to zero, which we take in this context to be the MIC, increases with inoculum size, $n_0$. Figure~\ref{fig:4N0}(b) plots this MIC as a function of $n_0$ from two sets of results: one which collates data from multiple assays done over a long period of time, and a second set where the data were collected in one experiment using the same peptide stock for each inoculum size. In both cases, the MIC approaches a constant value at low $n_0$, and increases sub-linearly at higher $n_0$ (the dashed line in Figure~\ref{fig:4N0}(b) has unit slope). 

Perhaps the simplest way to account for an inoculum effect, given our observation of depletion of active peptides by cell lysis, is that each lysed cell inactivates a fixed number of antimicrobial molecules. However, in its simplest form, this mechanism should lead to a linear dependence of the MIC on $n_0$ at high enough $n_0$. It is possible that our observed sub-linear $n_0$ dependence is {\it en route} to such a linear dependence, but we do not reach high enough inoculum concentrations to observe it. On the other hand, 
a speculative  explanation for a sub-linear dependence at all $n_0$ is that cells aggregate at high densities and high peptide concentrations,\footnote{We have seen such aggregation in optical microscopy (data not show).} which may \citep{Moiset2013} allow peptide molecules or complexes to affect two cells simultaneously.

\section{Discussion: What really {\it is} the MIC?}

The overwhelming majority of existing MIC measurements rely on a single series of dilution assays performed at one inoculum size (typically $5\times 10^5\,\si{\mbox{cells}\per\milli\litre}$) to determine a one-off concentration of antimicrobial agent at which no growth is observed after 24 hours. This `one-off MIC' is frequently assumed to represent, in the words of \cite{Melo2009}, `the macroscopically observable threshold for the onset of [antimicrobial] activity'. This implies that sub-MIC concentrations of antimicrobial agents are sub-lethal towards the target cells. In the light of our findings, both the practice of determining a `one-off MIC' value and the quoted interpretation of the significance of this value are problematic, at least for pexiganan acting on \ecoli.

First, we have found, using multiple replicates, that the `one-off MIC' shows large intrinsic variability, Figure~\ref{fig:96well}(a). This is because at these peptide concentrations, cell numbers drop very low before net regrowth starts, Figure~\ref{fig:KC_MHB}, and such low numbers give rise to significant stocasticity.  Instead of a `one-off MIC' from a single dilution series, findings from multiple replicates, reported as the fraction of samples in which no growth is observed after 24 hours as a function of the peptide concentration, Figure~\ref{fig:96well}(b), are necessary to give adequate information. 

Secondly, we have shown that exposure to sub-MIC levels of pexiganan kills a fraction of \ecoli\ in a population quickly, leaving the remaining cells to grow at a rate that is indistinguishable from that of cells never exposed to the AMP, Figure~\ref{fig:SC}. Indeed, the MIC for a standard inoculum size ($5\times10 ^{5}$ cell/ml) is $\sim 10\times$ above its single-cell action threshold: compare the black and red data sets in Figure~\ref{fig:4N0}(a). Thus, it is {\it not} true that the MIC of pexiganan is `the macroscopically observable threshold for the onset of [antimicrobial] activity'. Instead, if a single value is desired to quantify the `single-cell action threshold' of pexiganan on \ecoli\ MG1655, a reasonable candidate is the concentration for the onset of lengthening detection times, $\tau_{0.12}$: compare Figure~\ref{fig:subMIC}(d) with the red curve in Figure~\ref{fig:4N0}(a). 

The interpretation that the MIC itself is a `single-cell action threshold' clearly lies behind a substantial body of literature. Thus, \emph{e.g.}, \cite{Ramamoorthy2006} found a MIC of  \SI{4}{\micro\mbox{M}} for pexiganan acting on \ecoli\ using an inoculum of $10^7$ cell/ml, but observed outer membrane perturbation (according to an ANS uptake assay) at the significantly lower concentrations of \SI{0.67}{\micro\mbox{M}} after only 5 minutes of incubation. \cite{Ramamoorthy2006} infer that factors other than membrane disruption must be involved in the bacterial killing process. A later study came to a similar conclusion on similar grounds \citep{Pius2012}. Our findings mean that this inference is neither necessary nor likely correct: a fraction of cells should be killed (presumably by mechanisms involving membrane perturbation) at $\SI{0.67}{\micro\mbox{M}} \lesssim 0.2 \times \mbox{MIC}$. In general, then,  understanding what MIC values mean is crucial if pharmacodynamic results are to be correctly compared with mechanistic studies. 

Thirdly, there is strong evidence that peptides causing cell death are sequestered and unavailable for further bactericidal action.  This finding should influence the way in which comparative data between different AMPs are interpreted. Thus, \emph{e.g.}, Fluorogainin-1 is designed to be a `more stable' analogue of pexiganan. It displays a lower MIC against \emph{S. aureus} \citep{Gottler2008using}. Our findings suggest that this could be because Fluorogainin-1 is not depleted as rapidly. Separately, one could postulate that mutant strains more resistant to pexiganan \citep{Perron2006} could have evolved more efficient mechanisms for peptide inactivation in lysed cells, thus raising the measured MIC. These scenarios are speculative, but illustrate once more the need to consider carefully the meaning of MIC measurements.

Finally, we have quantified the inoculum effect (IE) of pexiganan acting on \ecoli, Figure~\ref{fig:4N0}(b). There is no IE for $n_0 \lesssim 10^3$~cells/ml; thereafter, $\mbox{MIC} \propto n_0^\alpha$ with $\alpha \approx 0.4$. This finding has significant practical and theoretical implications. 

Practically, while the IE is well known, it is seldom quantified. In lieu of quantification, it is presumably natural to assume that the MIC is essentially proportional to $n_0$, especially if the antimicrobial agent is known to be depleted by the process of killing bacteria. Making this assumption in the case of pexiganan acting on \ecoli\ MG1655 would lead to very substantial errors, with knock-on implication for dosage regimes arising from pharmacokinetic studies based on such an assumption. Theoretically, we note that our power law with $\alpha \approx 0.4$ also fits the data for 5 out of the 6 antibiotics for which the IE has been quantified \citep{Udekwu2009} (we find $\alpha$ between 0.3 and 0.5 fitting their data to power laws). Intriguingly, \cite{Udekwu2009} found no depletion of the antibiotic for 3 of these 5 antibiotics. It is a challenge for future theoretical modelling to decide whether a deeper generic mechanism underly such similarity, or whether two or more mechanisms fortuitously lead to similar quantitative IEs.

Overall, we suggest that data like those presented in Figure~\ref{fig:4N0}(a) should form the minimum basis for interpreting MIC measurements in the context of mechanistic studies, pharmacodynamic modelling, and clinical decision making for all antimicrobial agents. More specifically, given that many bactericidal AMPs are structurally similar to pexiganan, we surmise that our findings concerning the {\it modus operandi} of pexiganan, e.g., that its action is likely crucially linked to phenotypic heterogeneity of the target organism, may generalise to other compounds. For example, preliminary data on the MIC of amhelin, a pore-forming peptide designed by \cite{Rakowska2013}, shows the same growth curve (Figure \ref{fig:subMIC}) and IE (Figure \ref{fig:4N0}(b)) as pexiganan. Future research should explore the validity of this surmise for a wider range of AMPs, especially those that, like pexiganan, are bactericidal. 

\subsubsection*{Acknowledgements} AKJ was funded by an EPSRC studentship. JSL was funded by the National Physical Laboratory and SUPA. LR and MGR were funded by the UK Department of Business, Innovation and Skills. WCKP was funded by EPSRC Programme Grant EP/J007404/1. We thank Simon Titmuss for illuminating discussions, Angela Dawson for assistance in biological lab work and Vincent Martinez for assistance with data analysis.

\bibliographystyle{spbasic}
\bibliography{AMP}

\end{document}